\documentclass[a4paper,fleqn]{cas-sc}
\usepackage[numbers,sort&compress]{natbib}
\usepackage{amsmath,amssymb,booktabs,placeins}
\newcommand{\Lc}{L_c}
\newcommand{\PS}{\mathrm{PS}}
\newcommand{\Ssurf}{\mathrm{SS}}
\ExplSyntaxOn
\keys_set:nn { stm / mktitle } { nologo = true }
\ExplSyntaxOff
\begin{document}
\let\WriteBookmarks\relax
\let\printorcid\relax
\shorttitle{PINN-based pressure-driven blade optimization}
\shortauthors{Yuan Huang and Francesca di Mare}
\title[mode=title]{A physics-informed neural network method for pressure-driven turbine-blade optimization}
\author[1]{Yuan Huang}
\author[1]{Francesca di Mare}
\cormark[1]
\ead{francesca.dimare@ruhr-uni-bochum.de}
\affiliation[1]{organization={Chair of Thermal Turbomachines and Aero Engines, Ruhr University Bochum}, addressline={Universit\"atsstrasse 150}, city={Bochum}, postcode={44801}, country={Germany}}
\cortext[1]{Corresponding author.}
\begin{abstract}
We develop a physics-informed method for aerodynamic shape optimisation of turbine blades, treating the flow field and blade geometry as coupled unknowns. The method translates prescribed surface-pressure distributions into geometric modifications through pressure-target continuation and alternating flow--geometry updates. A differentiable representation of the blade surfaces allows their positions and orientations to be optimised within the current flow field, followed by flow re-adaptation to the modified boundaries. The inverse formulation uses the two-dimensional steady compressible Euler equations as physics constraints and is evaluated on the LS89 turbine cascade using 28 pressure-target points derived from experimental data. Starting from a modified profile and a warm-started flow network, the final pressure-side and suction-side root-mean-square fitting errors are 1.13\% and 2.21\% of inlet total pressure, respectively. Independent forward verification with SharC reproduces the principal surface-pressure trends of the experimental measurements, with remaining differences near the trailing edge. These results demonstrate the translation of experimental-origin loading data into geometric blade modifications through coordinated optimisation of the flow field and moving boundaries.
\end{abstract}
\maketitle

\section{Introduction}
\label{sec:introduction}

Turbine aerodynamic design proceeds through a hierarchy of increasingly detailed descriptions. Cycle and duty requirements establish quantities such as mass flow, pressure ratio, rotational speed, and required work. Preliminary mean-line design uses these requirements to determine stage count, velocity triangles, and principal dimensions, while through-flow and blade-row analyses refine the flow distribution before detailed blade design and verification \cite{aungier2006,agromayor2019,li2017turbomachinery}. At the blade level, geometric changes must realize the required aerodynamic loading within constraints on passage dimensions, thickness, and operating conditions. This step remains a central task in turbomachinery optimization because local changes to the surface can affect the flow throughout the passage \cite{li2017turbomachinery,zou2024turbomachinery}.

Surface pressure provides a direct connection between blade geometry and aerodynamic loading. The pressure difference between the two surfaces determines the pressure contribution to blade forces, while the chordwise pressure distribution describes the acceleration and diffusion experienced by the flow \cite{aungier2006}. Studies of highly loaded low-pressure turbines show that the placement of loading influences boundary-layer development, separation, and losses, with the outcome depending on Reynolds number, incidence, and incoming-flow conditions \cite{howell2001,berrino2016}. Pressure-based inverse design offers a way to prescribe this distribution and seek a geometry that realizes it; Daneshkhah and Ghaly demonstrate such a formulation for subsonic and transonic viscous compressor and turbine flows \cite{bladeinverse}. The engineering value of pressure matching therefore depends on selecting an appropriate loading target and subsequently assessing the resulting flow.

These design requirements also arise in thermal-energy-storage applications. Carnot batteries store electrical energy in thermal form and recover electricity through a power-conversion process, with component requirements that depend on the thermodynamic architecture \cite{dumont2020,liang2022}. For pumped thermal electricity storage, McTigue et al. show how component performance and operating conditions influence cycle performance \cite{mctigue2015}. The coupling between thermal stores and conversion machinery further makes departures from nominal operation relevant to system design, as examined in studies of decoupled thermal stores and off-design operation \cite{davenne2020,mctigue2024offdesign}. Such applications motivate blade-design tools that can accommodate different operating specifications and desired loading distributions. A calculation at a prescribed operating point is one component-level step toward that broader requirement; establishing performance over an operating envelope additionally requires calculations at the relevant conditions.

The problem considered here is to modify an initial turbine blade so that it produces a prescribed surface-pressure distribution under specified inlet and outlet conditions. The initial geometry supplies the starting design, the boundary conditions define the operating point, and the target pressures specify the desired aerodynamic response. The unknowns are the modified blade surface and the flow around it. Their coupling is essential: moving the surface changes the domain and its boundary conditions, while the resulting flow determines whether the pressure specification has been achieved.

Conventional aerodynamic shape optimization couples a parameterized geometry to a computational fluid dynamics (CFD) solver and updates the design using the computed response. Derivative-free methods, including evolutionary and population-based searches, avoid flow sensitivities but require repeated flow evaluations. Gradient-based methods use finite differences, direct sensitivities, or adjoint equations to guide the updates \cite{li2017turbomachinery}. For a scalar objective and many design variables, an adjoint calculation obtains sensitivities without a separate perturbed flow solution for every variable \cite{jameson1988aerodynamic,giles2000}. Within this CFD-based framework, pressure-driven inverse design specifies the desired surface-pressure or loading distribution directly, and geometric updates seek to reduce the discrepancy between computed and prescribed pressures \cite{giles1987two,dang1995euler}. Representative formulations include the viscous blade-design method of Daneshkhah and Ghaly and the control-theoretic turbine-cascade optimization of Feng et al. \cite{bladeinverse,controlinverse}. Pressure matching defines the objective; derivative-free and gradient-based algorithms provide ways to solve the resulting optimization problem.

A prescribed pressure distribution need not identify a unique admissible blade, particularly when it is available only at a finite set of surface locations. Inverse-problem formulations address weak identifiability and sensitivity to imperfect data through additional information and regularization \cite{regularization}; Bayesian approaches also quantify uncertainty in the unknowns through a posterior distribution \cite{stuart2010}. In deterministic blade redesign, geometric bounds, smoothness penalties, and passage constraints restrict the search to usable profiles. An initial blade can incorporate preliminary-design knowledge while permitting subsequent changes to both camber and thickness. These ingredients define the admissible design space and the operating specification against which the pressure target is evaluated.

Machine learning has been incorporated into aerodynamic optimization to reduce repeated evaluation costs and exploit information from previous calculations. Surrogate models approximate the relation between geometry, operating conditions, and aerodynamic outputs, while learned representations can support flow reconstruction and design updates \cite{li2022,zou2024turbomachinery}. Their usefulness depends on the training coverage, approximation accuracy, and computational cost of constructing the model. Physics-informed learning incorporates governing equations and boundary conditions into model training, allowing physical constraints to contribute information beyond labelled flow data \cite{piml}. This creates opportunities for design calculations in which the physical state is learned as part of the optimization, while making the accuracy of that state a central requirement.

Physics-informed neural networks (PINNs) represent physical fields using differentiable neural functions trained against differential equations, boundary conditions, and observations. Raissi, Perdikaris, and Karniadakis establish this framework for forward and inverse problems involving nonlinear partial differential equations, and Karniadakis et al. review its place within physics-informed machine learning \cite{pinn,piml}. Automatic differentiation supplies spatial derivatives for the equation residuals and derivatives with respect to trainable physical or geometric quantities. This permits a field representation and design variables to participate in the same optimization without requiring a precomputed geometry-to-flow database for every formulation. Obtaining a physically accurate state during optimization remains necessary, however, and depends on how the governing equations and boundary conditions are satisfied.

Relevant fluid applications include the inference of velocity and pressure from incomplete flow observations in hidden fluid mechanics \cite{hidden}, the range of forward and inverse problems reviewed by Cai et al. \cite{cai2021}, and PINN solution techniques for steady internal Euler flows studied by Post, Winhart, and di Mare \cite{internal}. Compressible-flow applications also include forward and inverse high-speed Euler problems and supersonic inverse problems involving shocks and expansion waves \cite{mao2020highspeed,jagtap2022supersonic}. These works establish ways to combine differentiable flow representations with physical constraints. In blade redesign, the wall location and orientation are additional unknowns, and both pressure matching and the wall condition must be evaluated on the evolving surface.

Geometry has been incorporated into physics-informed models through several distinct representations. Coordinate-mapping approaches, such as PhyGeoNet, also accommodate irregular domains in physics-constrained flow learning \cite{gao2021phygeonet}. Oldenburg et al. encode irregular vessel shapes with a variational autoencoder, concatenate the resulting latent description with the spatial coordinates supplied to a PINN, and use a separate boundary-constraining network to impose boundary conditions \cite{oldenburg2022}. This construction produces a surrogate over a family of non-parameterized geometries without requiring CFD flow labels. Its geometry encoder requires a pre-generated shape ensemble, and the encoded geometry is a conditional input to the flow surrogate rather than an unknown boundary displaced by an aerodynamic objective. The approach therefore addresses prediction across a represented geometry family, whereas pressure-driven redesign requires the geometry itself to be determined from the prescribed loading.

Cao, Song, and Zhang address another geometric difficulty by mapping a body-fitted airfoil mesh to a uniform computational domain and training the PINN in the transformed coordinates \cite{cao2024}. The transformation removes the airfoil as an internal boundary of the learning domain and redistributes resolution near the leading edge, where a PINN in physical coordinates has difficulty representing rapid flow variations. Their study establishes a forward solver for inviscid subsonic airfoil flows and demonstrates parametric prediction with respect to incidence. It requires a suitable mesh transformation for the geometry being analyzed and is not formulated as coupled pressure-driven shape optimization. Thus, the mapping regularizes flow learning on a prescribed airfoil, whereas the present problem additionally requires the aerodynamic boundary to move in response to a pressure objective.

Wong et al. take a hybrid route: their low-fidelity-influenced PINN uses flow fields from a viscous--inviscid coupled airfoil-analysis tool together with residuals of the steady incompressible two-dimensional Navier--Stokes equations, and the resulting surrogate supports inverse design using target pressure-coefficient distributions \cite{wong2024}. This establishes a route in which a physics-informed aerodynamic surrogate supplies flow information to the inverse-design process, while retaining a dependence on the coverage and modeling assumptions of the external low-fidelity generator. Lu et al. directly optimize the physical state and a continuous material-density field together using penalty and augmented-Lagrangian formulations of hard-constrained PINNs \cite{hpinn}. Their Stokes-flow example represents topology through density-dependent permeability on a fixed domain. This formulation provides a precedent for jointly optimizing physical states and design variables, while representing geometry through a distributed material field rather than an explicitly moving wall. Related PDE-constrained optimization studies address implicit state sensitivities through bi-level PINNs \cite{hao2023bilevel} and the balance between design objectives and physical losses \cite{mowlavi2023optimal}.

Sun, Sengupta, and Juniper directly address airfoil shape optimization using a geometry-parameterized PINN \cite{sun2023pinnairfoil}. Their network takes spatial coordinates and design parameters as inputs, coupling surrogate training with lift-to-drag optimization through automatic differentiation. The present formulation instead updates explicit blade-surface control variables within a spatial flow field that is repeatedly adapted to the current geometry. This yields a different coupling between the learned state and the moving boundary, with pressure matching as the design objective.

Together, these studies establish geometry-aware flow prediction and physics-informed inverse design as relevant precedents \cite{oldenburg2022,cao2024,wong2024,hpinn,sun2023pinnairfoil}. The present work focuses on a pressure-driven formulation in which the flow-network parameters and explicit blade-surface control variables are both updated during redesign. In this formulation, pressure mismatch can decrease either through adaptation of the flow field or through displacement of the wall within that field. Once the wall moves, the preceding flow is no longer guaranteed to satisfy the boundary conditions on the modified geometry. This motivates an update strategy that coordinates boundary motion with subsequent flow re-adaptation.

Two aspects of the coupled optimization are therefore considered: the coordination of flow and geometry updates at a given pressure target, and the progression of that target from the initialized loading to the desired distribution.

Curriculum learning organizes training through a progression of tasks \cite{bengio2009}. This strategy is relevant to PINNs, whose non-convex objectives combine equation, boundary-condition, and data residuals with potentially unbalanced gradients and unequal convergence rates \cite{gradient,ntk}. A continuation-based curriculum introduces intermediate problems and reuses previously learned states to guide successive adaptation. Krishnapriyan et al., for example, progressively alter physical coefficients while initializing each stage with the preceding trained network \cite{krishnapriyan2021}. The aim is to facilitate optimization through incremental changes, without assuming that the objective becomes convex or that global convergence is guaranteed.

Related progressive strategies have been developed under different formulations and terminology. For hemodynamic modeling, Chen et al. introduce the Boundary Progressive PINN, which begins with simplified vascular boundaries and progressively incorporates geometric details \cite{chen2025boundary}. For stationary compressible Euler flows, Wassing et al. first train with a prescribed artificial-viscosity reference and then reduce that reference during continued optimization, while allowing the network to retain locally adaptive dissipation where needed \cite{wassing2024}. These examples extend the progression from physical coefficients to boundary complexity and artificial dissipation.

Other formulations reuse earlier predictions or explicitly track changes in problem parameters. Howard et al. combine stacked physics-informed networks with progressive changes in PDE parameters; their physics-informed DeepONet example for Burgers' equation decreases viscosity from $10^{-2}$ to $10^{-3}$ and then $10^{-4}$, while passing earlier predictions to subsequent models as low-fidelity information \cite{howard2025}. Chen et al. develop homotopy dynamics for singularly perturbed PDEs, advancing neural approximations from larger perturbation parameters toward more demanding target regimes \cite{chen2025homotopy}. Zhu et al. combine viscosity continuation with adaptive lifting: a coarse higher-viscosity solution supplies an auxiliary coordinate representation, and the solution-network parameters are carried forward to lower-viscosity stages \cite{zhu2025lifting}. Although their continuation variables and mechanisms for reusing learned information differ, these approaches provide precedents for approaching a difficult target through a sequence of modified problems.

Our earlier progressive Euler-PINN framework used boundary-progressive training and network-weight reuse for forward flow prediction across prescribed blade geometries \cite{huang2026fast}. The present work builds on this differentiable flow representation and reuse of learned states to determine an unknown blade geometry from a prescribed pressure distribution. We interpolate between the initialized and desired surface pressures, carrying the flow--geometry state forward as the target advances. This continuation changes the pressure specification, rather than only loss weights, and requires no prescribed sequence of intermediate blade shapes. The three-stage alternating updates introduced here coordinate flow adaptation and boundary motion within each pressure-target step.

We therefore propose a pressure-homotopy alternating PINN algorithm with differentiable control-point representations of both blade surfaces. Each optimization cycle comprises three substeps: flow adaptation, a frozen-field geometry update, and flow re-anchoring. Because the current pressure target need not be compatible with a flow satisfying all boundary conditions on the fixed blade, the first substep adapts the flow with a relaxed wall penalty to construct an intermediate field that guides boundary motion. The second substep freezes the network parameters and updates the blade surfaces using pressure-matching, wall-condition, and geometric-regularization terms, with derivatives retained through the boundary coordinates and normals. The third substep fixes the modified geometry and updates the flow with renewed emphasis on the wall condition. This sequence coordinates pressure-error reduction through field adaptation and boundary displacement, with physical consistency assessed after each cycle.

The method is evaluated on a two-dimensional LS89 reference-based benchmark governed by the steady Euler equations, starting from a modified half-thickness profile and a warm-started neural flow field. The primary assessment is whether independent forward CFD on the redesigned blade reproduces the prescribed pressure distribution; the reference profile provides a visual shape comparison. At the prescribed final cycle, pressure-fitting errors are reported together with wall-condition errors and sampled equation residuals. Independent forward verification with SharC assesses whether the redesigned blade produces the prescribed loading when evaluated outside the inverse network. The benchmark therefore distinguishes pressure recovery from exact geometric reconstruction. Efficiency and loss assessments additionally require suitable loading targets and flow models capable of representing the relevant mechanisms \cite{aungier2006,howell2001,berrino2016}. The following sections present the formulation, algorithm, and the inverse, geometric, CFD, and experimental comparisons.

\section{Pressure-driven redesign problem and computational setting}
\subsection{Design objective}

Given an initial blade $\mathbf q_0$, operating conditions, and prescribed pressure samples $p^\star$, the inverse problem seeks blade variables $\mathbf q$ and neural-flow parameters $\boldsymbol\theta$. The primary design metric is the pressure mismatch on the evolving pressure and suction surfaces,
\begin{equation}
 J_p(\boldsymbol\theta,\mathbf q)=\frac{1}{2}\sum_{s\in\{\PS,\Ssurf\}}\frac{1}{N_s}\sum_{i=1}^{N_s}\left[\frac{p_{\boldsymbol\theta}(x_{s,i},y_s(x_{s,i};\mathbf q))-p^\star_{s,i}}{p_{t,\mathrm{in}}}\right]^2.
 \label{eq:redesign_objective}
\end{equation}

Pressure is nondimensionalized by inlet total pressure. Equation~\eqref{eq:redesign_objective} is optimized together with sampled flow-equation, boundary, wall, and geometry penalties. The LS89 reference coordinates do not enter these objectives and are used only after training for a benchmark geometry diagnostic.
\subsection{Geometry, pressure data, and training information}

The domain $\Omega(\mathbf q)$ is a periodic two-dimensional cascade passage. Ten equally spaced streamwise stations, including the leading and trailing edges, are fixed on each surface; the eight interior ordinates on each side are trainable, giving $\mathbf q\in\mathbb R^{16}$. The endpoints, pitch, and outer-domain topology remain fixed. In the adopted coordinate system, $x_{\rm LE}/\Lc=0$, $x_{\rm TE}/\Lc=0.54630$, the pitch is $0.850\Lc$, and the inlet and outlet are at $x/\Lc=-0.22156$ and $0.88625$. The downstream periodic extensions have slope $-1.23237$.

The initial blade is constructed by retaining the camber line and reducing the reference thickness by half. With camber $c=(y_\PS+y_\Ssurf)/2$ and half-thickness $h=(y_\Ssurf-y_\PS)/2$, the initial surfaces are $y_{\PS,0}=c-0.5h$ and $y_{\Ssurf,0}=c+0.5h$. This construction reflects a redesign setting in which preliminary mean-line design has already established the operating conditions and the initial camber distribution is retained. SharC is used to compute the flow on this reduced-thickness geometry. A network with the same architecture as the inverse model is first trained on these CFD flow data, and its parameters initialize the subsequent pressure-driven optimisation. This preliminary fit provides an approximate flow field and a physically informed starting point; a highly accurate forward solution from the network is unnecessary because flow adaptation and re-anchoring continue throughout the inverse calculation.

The fixed-$x$ target file contains eight pressure-side and twenty suction-side records. The source curve was digitized from an experimental plot using axis calibrations $s/\Lc\in[0,1.4]$ and $p/p_t\in[1,0.2]$ \cite{internal}; the pressure-side leading-edge value $p/p_t=1$ was added as a synthetic stagnation condition. The design target comprises 27 digitized experimental values and one synthetic leading-edge condition.

For surface $s\in\{\PS,\Ssurf\}$, each pressure sample has a fixed streamwise station $x_{s,i}$ and target $p^\star_{s,i}$. Its transverse coordinate follows the wall:
\begin{equation}
 \mathbf z_{s,i}(\mathbf q)=[x_{s,i},y_s(x_{s,i};\mathbf q)]^{\mathsf T}.
 \label{eq:sensor}
\end{equation}

The observation abscissae remain fixed while their ordinates follow the current wall. Stored raw surface-coordinate labels, rather than remapped current arc length, provide the denominators for the finite-difference pressure-gradient target.

\subsection{Flow model and boundary conditions}

At the prescribed design operating point, viscous effects are expected to have a limited influence on the main surface-pressure distribution. The compressible Euler equations are therefore adopted for the inverse PINN, on the assumption that their predicted pressure loading is close to that of a compressible Navier--Stokes model under these conditions. Viscous SharC calculations supply the preliminary flow data and the independent forward verification, allowing the pressure response of the redesigned blade to be checked with viscous effects included. The results in Section~\ref{sec:forward_results} support this modelling choice for the present case: although the inverse calculation uses Euler physics and experimental-origin pressure targets, the resulting blade reproduces the principal measured pressure-loading trends when evaluated with viscous SharC. The redesigned geometry thus achieves good agreement with the prescribed pressure distribution beyond the inviscid inverse model. This modelling approximation concerns surface-pressure loading; viscous losses and near-wall transport require the viscous formulation.

The fluid is calorically perfect with $\gamma=1.4$. Coordinates are normalized by chord and pressure by inlet total pressure. The nondimensional thermodynamic relations are $T=\gamma p/\rho$ and $T_t=T+(\gamma-1)(u^2+v^2)/2$, with inlet $p_t=T_t=1$. The static outlet pressure is prescribed for $M_{\rm out}=0.84$ as
\begin{equation}
 p_{\rm out}=[1+(\gamma-1)M_{\rm out}^2/2]^{-\gamma/(\gamma-1)}.
\end{equation}

At inlet points, $T_i=1-0.2(u^2+v^2)$, $p_i=T_i^{3.5}$, and $\rho_i=1.4p_i/T_i$. The inlet loss is the sum of the mean squared errors in $\rho-\rho_i$, $p-p_i$, $u-|u|$, and $v$; the outlet loss penalizes $p-p_{\rm out}$ and $u-|u|$. For each upstream and downstream pair, the periodic loss is the mean over points and all four components of the squared $(\rho,u,v,p)$ state difference. The total-temperature term is the sum of the domain mean and outlet mean of $(T_t-1)^2$. An additional mean inlet axial-velocity constraint, $\bar u_{\rm ref}\approx0.131$, is obtained from the reference CFD solution for the LS89 reference blade. The term $\mathcal L_{\bar u}=(\operatorname{mean}_{\rm in}u-\bar u_{\rm ref})^2$ has weight 100 and supplies flow information beyond the surface-pressure targets.

The flow network $f_{\boldsymbol\theta}(x,y)=(\rho,u,v,p)$ has three hidden layers of 20 $\tanh$ units; exponential maps enforce positive density and pressure. Geometry enters through physical coordinates, not through a separate feature channel. The primitive steady-Euler residual is
\begin{align}
 R_\rho&=u\rho_x+v\rho_y+\rho(u_x+v_y),\\
 R_u&=\rho(uu_x+vu_y)+p_x,\\
 R_v&=\rho(uv_x+vv_y)+p_y,\\
 R_p&=up_x+vp_y+\gamma p(u_x+v_y).
 \label{eq:euler}
\end{align}

Spatial derivatives are obtained by automatic differentiation, and $\mathcal L_E$ is the mean squared residual over all four components and all sampled domain points. This residual formulation is applied to smooth, steady inviscid flow.

\section{Pressure-homotopy alternating redesign algorithm}

\begin{figure}[pos=htbp]
\centering
\includegraphics[width=\linewidth]{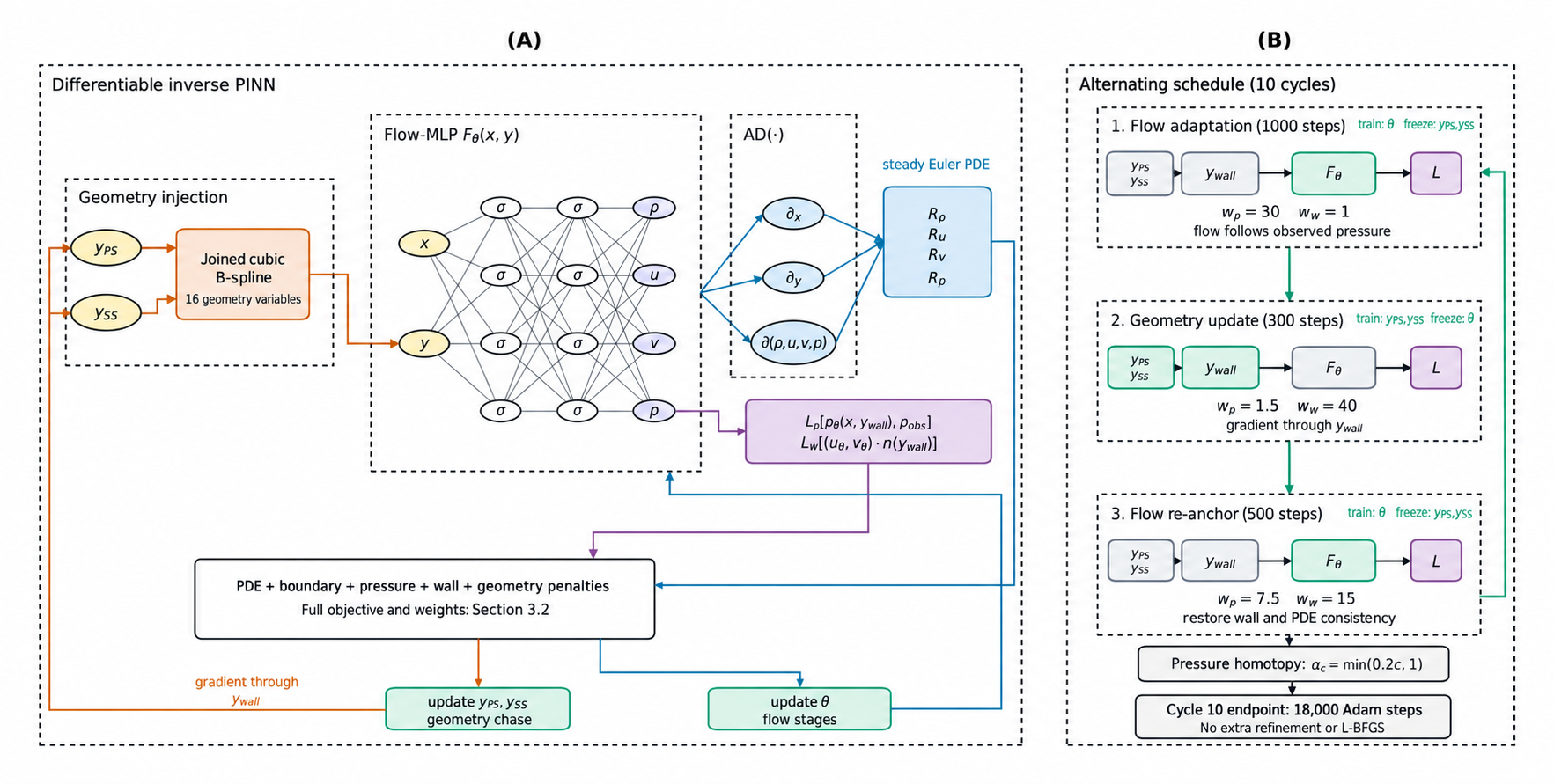}
\caption{Differentiable inverse PINN architecture and the exact ten-cycle optimization schedule. (A) Schematic neural architecture with geometry injection through the joined cubic B-spline wall representation, automatic-differentiation paths for the steady Euler residuals and wall/pressure objectives, and separate geometry and flow updates; the network sketch is schematic (actual network: 3 hidden layers of width 20). (B) Each cycle performs flow adaptation (1000 steps; $w_p=30$, $w_w=1$), a frozen-field geometry update (300 steps; $w_p=1.5$, $w_w=40$), and flow re-anchoring (500 steps; $w_p=7.5$, $w_w=15$). The current homotopy target $\alpha_c=\min(0.2c,1)$ is set at the start of each cycle. The ten cycles comprise 18,000 Adam steps.}
\label{fig:architecture}
\end{figure}

Figure~\ref{fig:architecture} summarizes the differentiable inverse formulation and the exact sequence executed in each of ten cycles.
\subsection{Differentiable wall representation}

The joined cubic B-spline represents one contour traversed from the pressure-side trailing edge through the leading edge to the suction-side trailing edge. Let $t\in[0,1]$ and $x_{\rm TE}=0.54630\Lc$. The streamwise coordinate is prescribed exactly as
\begin{equation}
 x(t)=x_{\rm TE}(2t-1)^2,
 \label{eq:joined_x}
\end{equation}
with $t_\PS(x)=[1-\sqrt{x/x_{\rm TE}}]/2$ and $t_\Ssurf(x)=[1+\sqrt{x/x_{\rm TE}}]/2$. The joined ordinate data use the ten pressure-side stations in reverse order and the suction-side stations translated by one pitch, omitting the duplicated leading-edge point. A cubic interpolant $y(t)=\sum_j a_jB_j(t)$ is determined from $A_{ij}=B_j(t_i)$ and $\mathbf a=A^{-1}\mathbf Y$. The knot vector repeats each endpoint four times and assigns each internal knot the mean of an adjacent triple. Local passage coordinates are $y_\PS(x)=y[t_\PS(x)]$ and $y_\Ssurf(x)=y[t_\Ssurf(x)]-0.85\Lc$.

The parametric contour is smooth through $t=0.5$, where the two surfaces share the fixed leading-edge position and parametric derivatives. At the vertical nose, $\mathrm dx/\mathrm dt=0$, so smoothness is defined for the parametric contour rather than the graph functions $y_s(x)$. Away from the nose, $\mathrm dy/\mathrm dx=(\mathrm dy/\mathrm dt)/(\mathrm dx/\mathrm dt)$. The graph normal is $\mathbf n_s=(-y'_s,1)/\sqrt{1+(y'_s)^2}$; its sign is immaterial to the squared wall penalties.

\subsection{Pressure homotopy and core objective}

Let $p^{(0)}_{s,i}$ denote the initial-field surface pressure interpolated to the observation stations. The cycle target is
\begin{equation}
 p^{(c)}_{s,i}=(1-\alpha_c)p^{(0)}_{s,i}+\alpha_cp^\star_{s,i},\qquad
 \alpha_c=\min(0.2c,1),\quad c=1,\ldots,10.
\end{equation}
Cycles 1--5 therefore replace one full change from $p^{(0)}$ to $p^\star$ by five increments of $0.2(p^\star-p^{(0)})$; Cycles 6--10 retain the full target. This schedule follows the curriculum-learning and continuation rationale discussed in the Introduction: intermediate problems reuse the preceding learned state to initialize subsequent adaptation \cite{bengio2009,krishnapriyan2021,chen2025boundary,wassing2024}. Here, carrying both flow and geometry forward limits the pressure change presented to each early cycle and is intended to facilitate optimization of the coupled non-convex problem. The continuation acts on the prescribed pressure values themselves. Both surfaces share $\alpha_c$, which specifies common scheduling rather than geometric symmetry.

The surface-balanced pressure loss at cycle $c$ is
\begin{equation}
 \mathcal L_p^{(c)}=\sum_s\frac{1}{N_s}\sum_i
 [p_{\boldsymbol\theta}(\mathbf z_{s,i})-p^{(c)}_{s,i}]^2.
\end{equation}
The normalization gives each surface equal aggregate importance despite unequal sample counts. The pressure term is the design driver; Euler and external-boundary terms constrain the flow, and wall terms couple that flow to the moving blade. The pressure-gradient term discourages local slope mismatch, while the gap, bounds, and ordinate-smoothness terms restrict the geometric update. With inlet, outlet, periodic, and total-temperature losses grouped as $\mathcal L_B$, the exact active objective is
\begin{align}
 \mathcal L^{(c,j)}={}&20\mathcal L_E+\mathcal L_B+100\mathcal L_{\bar u}
 +w_p^{(j)}\mathcal L_p^{(c)}+w_w^{(j)}\mathcal L_w+0.5\mathcal L_{\rm tan}\nonumber\\
 &+0.01\chi_j\mathcal L_{\nabla p}+100\mathcal L_{\rm gap}
 +10\mathcal L_{\rm bounds}+2\mathcal L_{\rm smooth},
\end{align}
where $j$ denotes the stage and $\chi_j=1$ only for the geometry update. The auxiliary gradient, wall-mask, normalization, gap, bounds, and ordinate-smoothness definitions are collected in Appendix~\ref{app:loss_definitions}; they stabilize or constrain the implemented update but do not define the pressure-homotopy path.

\subsection{Frozen-flow geometry update and flow re-anchoring}

Transpiration-based inverse design provides a more specific physical precedent. L\'{e}onard and Van den Braembussche used an intermediate Euler flow to guide blade modifications toward the prescribed loading \cite{leonard1992design}; Demeulenaere and Van den Braembussche extended the transpiration approach to three-dimensional blading \cite{demeulenaere1998three}. These methods motivate allowing an intermediate flow to guide boundary motion before re-solving the state. The present implementation retains the wall condition as a soft training penalty and moves the geometry using a frozen-field loss gradient.

The staged update is motivated by the state--design iteration used in CFD-based shape optimization: solve the flow on a given geometry, update the design, and solve the flow again on the modified geometry \cite{li2017turbomachinery,giles2000}. Our distinction is to embed this iteration within the pressure curriculum of a differentiable PINN, rather than use it only as an outer sequence of separate flow solves. The flow-adaptation stage updates the network parameters with the geometry fixed; the geometry stage freezes those parameters and propagates gradients through the current field's pressure, boundary coordinates, and normals to the B-spline controls; re-anchoring then updates the network on the modified wall. Thus, at each substage, the active gradient path is channeled through a prescribed parameter block instead of allowing simultaneous unconstrained descent over the combined flow--geometry variables. In this way, the method restructures the optimization by transforming a difficult, highly non-convex joint problem into multiple controlled stage subproblems, with the aim of facilitating optimisation through controlled flow--geometry coupling.

Each cycle follows the three-stage sequence in Fig.~\ref{fig:architecture}(B), with the updated parameters, step counts and weights listed in Table~\ref{tab:stages}; a new Adam optimizer is instantiated for every stage, and both flow stages use learning rate $10^{-5}$. First, flow adaptation holds $\mathbf q$ fixed and updates $\boldsymbol\theta$ toward the current homotopy target. Its wall weight is 1, lower than the later re-anchoring value of 15, so the field can accommodate the changed pressure specification before the boundary moves. The wall condition remains present, together with the tangency term; this stage relaxes its relative emphasis rather than removing it.
\begin{table}[pos=htbp]
\caption{Stages within every cycle; pressure weights refer to the current homotopy target.}\label{tab:stages}
\centering\begin{tabular}{lrrrr}\toprule
Stage & Updated parameters & Steps & $w_p$ & $w_w$\\\midrule
Flow update & $\boldsymbol\theta$ &1000&30&1\\
Geometry update &$\mathbf q_\PS,\mathbf q_\Ssurf$&300&1.5&40\\
Flow re-anchor &$\boldsymbol\theta$&500&7.5&15\\\bottomrule
\end{tabular}\end{table}

Second, the geometry stage freezes $\boldsymbol\theta$ but does not freeze the field's coordinate derivatives. At a fixed observation station, $\partial x/\partial\mathbf q=0$ and $\partial p/\partial\mathbf q=p_y\,\partial y/\partial\mathbf q$; changes in the wall normal are also differentiated. The implemented update is therefore
\begin{equation}
 \left.\frac{\partial\mathcal L}{\partial\mathbf q}\right|_{\boldsymbol\theta}=
 \frac{\partial\mathcal L}{\partial\mathbf z_s}\frac{\partial\mathbf z_s}{\partial\mathbf q}
 +\frac{\partial\mathcal L}{\partial\mathbf n_s}\frac{\partial\mathbf n_s}{\partial\mathbf q}
 +\frac{\partial\mathcal L_{\rm reg}}{\partial\mathbf q}.
\end{equation}
It moves the wall through pressure and wall-condition information sampled from the current field. This is a frozen-field partial derivative, distinct from a converged-flow total sensitivity or adjoint derivative,
\begin{equation}
 \frac{\mathrm dJ(\boldsymbol\theta^*(\mathbf q),\mathbf q)}{\mathrm d\mathbf q}
 =J_{\mathbf q}+J_{\boldsymbol\theta}\frac{\mathrm d\boldsymbol\theta^*}{\mathrm d\mathbf q},
\end{equation}
because the second term is absent. During each geometry-gradient evaluation, the interior collocation coordinates are treated as fixed with respect to $\mathbf q$. The interior Euler residual therefore supplies no direct geometry gradient, although it remains in the objective with weight 20. Its gradient with respect to the network parameters is active during flow adaptation and re-anchoring.

Third, flow re-anchoring fixes the modified $\mathbf q$ and resumes optimization of $\boldsymbol\theta$ with a wall weight of 15 and pressure weight of 7.5. The field used to guide the geometry step was trained on the preceding wall and is not automatically compatible with the displaced boundary. Re-anchoring therefore restores emphasis on the Euler and boundary losses on the new domain before the state is carried into the next homotopy cycle. Physical consistency after this finite, soft-constrained update is assessed using the reported flow and boundary residuals.

Geometry learning rates in Cycles 1--5 are $(1.8,1.6,1.4,1.2,1.0)\times10^{-4}$; later cycles use $10^{-4}$. The coordinatewise clipping radii are $(0.014,0.013,0.012,0.011,0.010)\Lc$, followed by $0.010\Lc$. Each radius is measured from the ordinate at the start of that geometry stage and does not bound continuous-surface displacement.

Training uses 3000 domain points and 1000 wall points per surface, resampled every 100 steps. Each boundary draw includes 50 inlet points, 50 outlet points, and 50 paired points on each upstream and downstream periodic segment. Ten cycles comprise 18000 optimizer steps and conclude with the Cycle 10 flow re-anchoring stage.

Results are reported after the final Cycle 10 re-anchoring stage. The reference profile provides a visual comparison with the final design.

\section{Evaluation protocol}
% Evaluation focuses on the pressure response of the redesigned geometry.
The redesigned blade is evaluated by whether an independent forward CFD calculation reproduces the prescribed surface-pressure distribution. The reference LS89 profile provides a visual comparison of the resulting shape. Pressure agreement within the inverse network is reported alongside this forward assessment.

Pressure error is
\begin{equation}
 E_{p,s}=100\sqrt{\frac{1}{N_s}\sum_i[(p_{\boldsymbol\theta}(\mathbf z_{s,i})-p^\star_{s,i})/p_{t,\rm in}]^2},\qquad
 \bar E_p=\tfrac12(E_{p,\PS}+E_{p,\Ssurf}).
\end{equation}

The reported mean is the arithmetic average of the pressure-side and suction-side RMSEs, giving both surfaces equal weight. Cycle comparisons use the same final target $p^\star$.

Independent forward verification uses SharC (Structured High-order Aerothermochemistry Real Fluid Computations), the in-house compressible-flow solver developed at Ruhr University Bochum. SharC uses a density-based finite-volume formulation on multi-block structured grids \cite{post2021condensing,internal}. Its applications include large-eddy simulation of condensing wet-steam flow in a turbine cascade \cite{post2021condensing} and reference CFD calculations for internal-flow studies \cite{internal}. The forward verification uses a viscous compressible Navier--Stokes formulation to assess the pressure response of the redesigned blade independently of the Euler-based inverse network.

% Compare independent CFD pressures at the available experimental stations.
For the forward assessment, pressures are normalized by the inlet total pressure and compared with the experimental measurements and published reference CFD. Position is measured along each blade surface from its leading edge.

\section{Results}
\subsection{Pressure fit and associated flow field}

\begin{figure}[pos=htbp]
\centering
\begin{minipage}[t]{0.28\linewidth}
\vspace{0pt}\centering
\includegraphics[width=\linewidth]{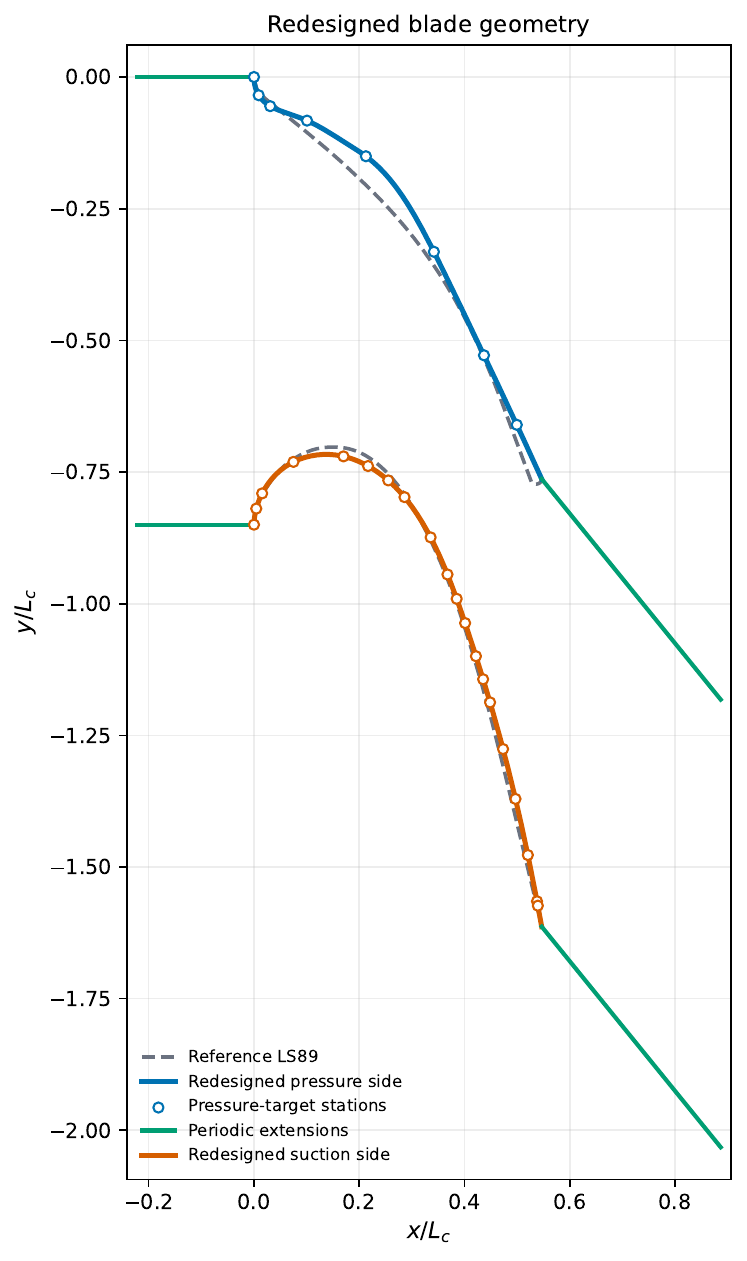}
\par\small (a) Redesigned blade geometry
\end{minipage}\hfill
\begin{minipage}[t]{0.70\linewidth}
\vspace{0pt}\centering
\includegraphics[width=\linewidth]{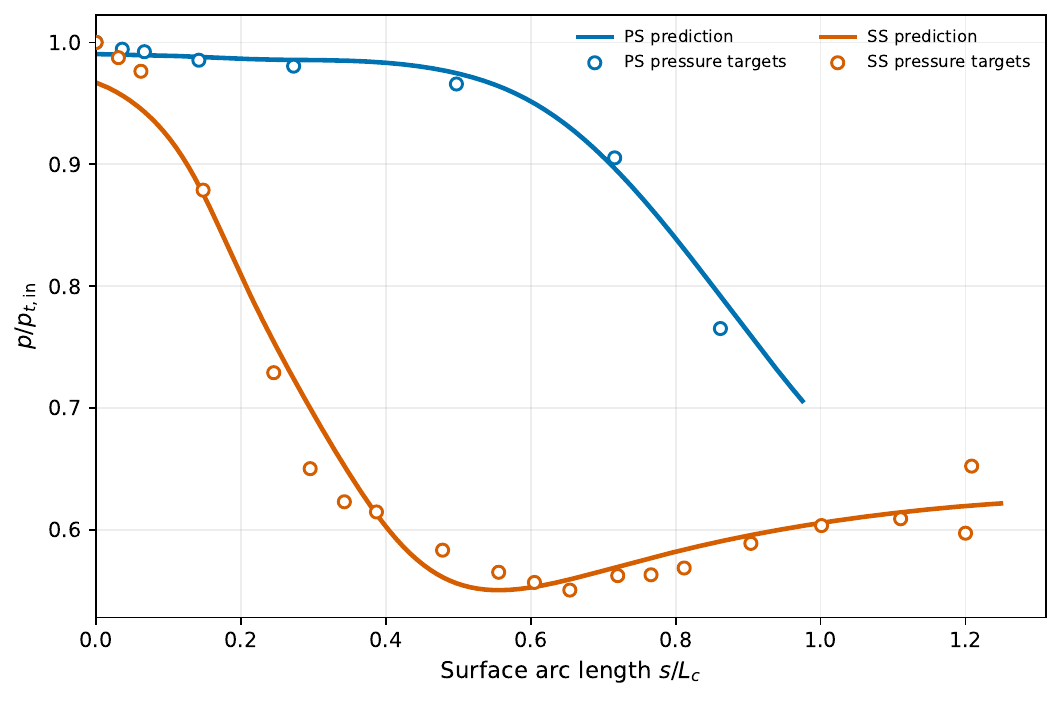}
\par\small (b) Inverse-PINN pressure prediction
\end{minipage}
\caption{Cycle 10 geometry and pressure fit from the final inverse-trained model. (a) Redesigned surfaces with the digitized reference LS89 outline. (b) Surface pressures predicted directly by the PINN (lines) and the prescribed pressure-target points (circles), plotted against arc length $s/\Lc$ along each redesigned surface from its leading edge. Independent forward-CFD pressures are reported in Fig.~\ref{fig:forward_cfd}.}\label{fig:pressure}\label{fig:geometry}
\end{figure}

At the prescribed Cycle 10 endpoint, pressure RMSE is 1.1279\% on the pressure side and 2.2086\% on the suction side, giving a surface-balanced mean of 1.6682\%. Figure~\ref{fig:pressure} presents the redesigned blade together with the surface pressures predicted directly by the inverse-trained PINN and the prescribed pressure-target points. The suction-side error is approximately twice the pressure-side error, indicating that the remaining pressure mismatch is distributed unevenly between the two surfaces. These values quantify agreement between the PINN output and the supplied pressure targets. Independent forward CFD subsequently tests the pressure response of the resulting geometry in Section~\ref{sec:forward_results}.
\FloatBarrier
The reconstructed suction surface closely follows the reference outline in Fig.~\ref{fig:geometry}(a), whereas a larger deviation remains in the middle of the pressure surface. Aerodynamic inverse design can be non-unique, depending on the prescribed data and geometric constraints; existence and uniqueness have been examined for turbine cascades, and different wing geometries have been reported to produce the same prescribed surface-pressure distribution \cite{pascoa2004existence,jeong1999inverse}. This possibility motivates assessing the redesigned blade through an independent forward CFD calculation rather than judging it solely by its resemblance to the reference outline.

\begin{figure}[pos=htbp]
\centering\includegraphics[width=0.94\linewidth,height=0.64\textheight,keepaspectratio]{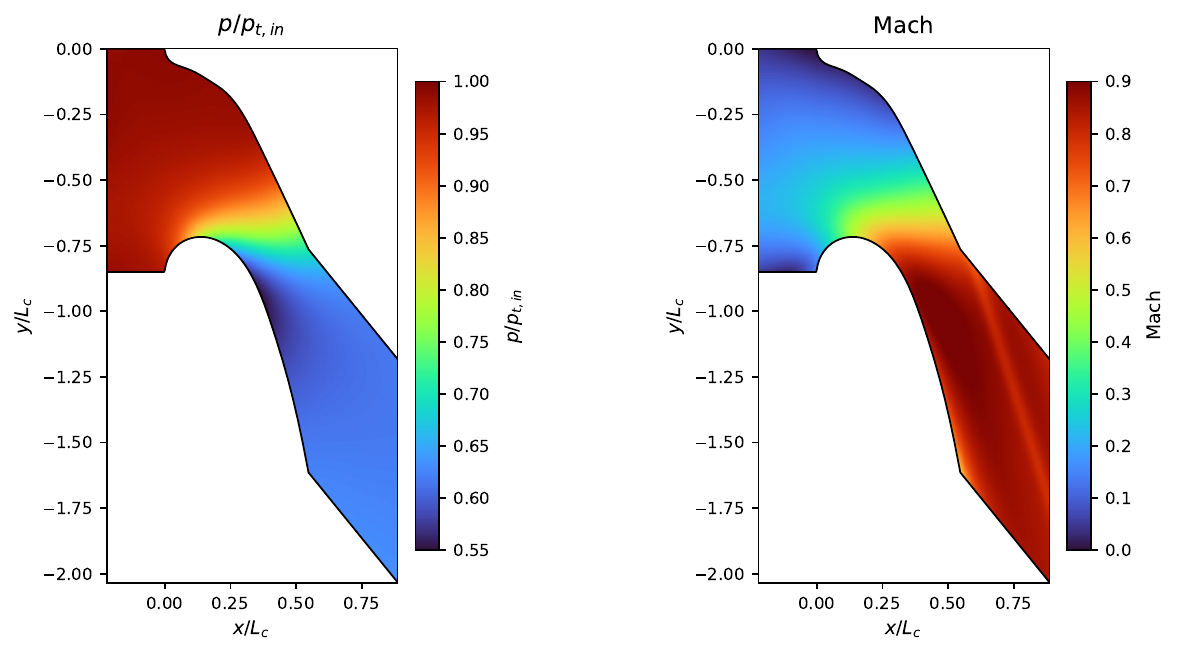}
\caption{Cycle 10 pressure (left) and Mach (right) fields from the same inverse-trained network as Fig.~\ref{fig:pressure}. Equation-residual and boundary diagnostics accompany these fields in Table~\ref{tab:physics}.}\label{fig:flow}
\end{figure}

Figure~\ref{fig:flow} shows the pressure and Mach fields predicted by the inverse-trained network at this endpoint. These fields provide the spatial context for the surface-pressure fit. Their equation residuals and boundary errors are evaluated in Table~\ref{tab:physics}.

\subsection{Pressure-ramp and full-target evolution}

\begin{figure}[pos=htbp]
\centering\includegraphics[width=\linewidth]{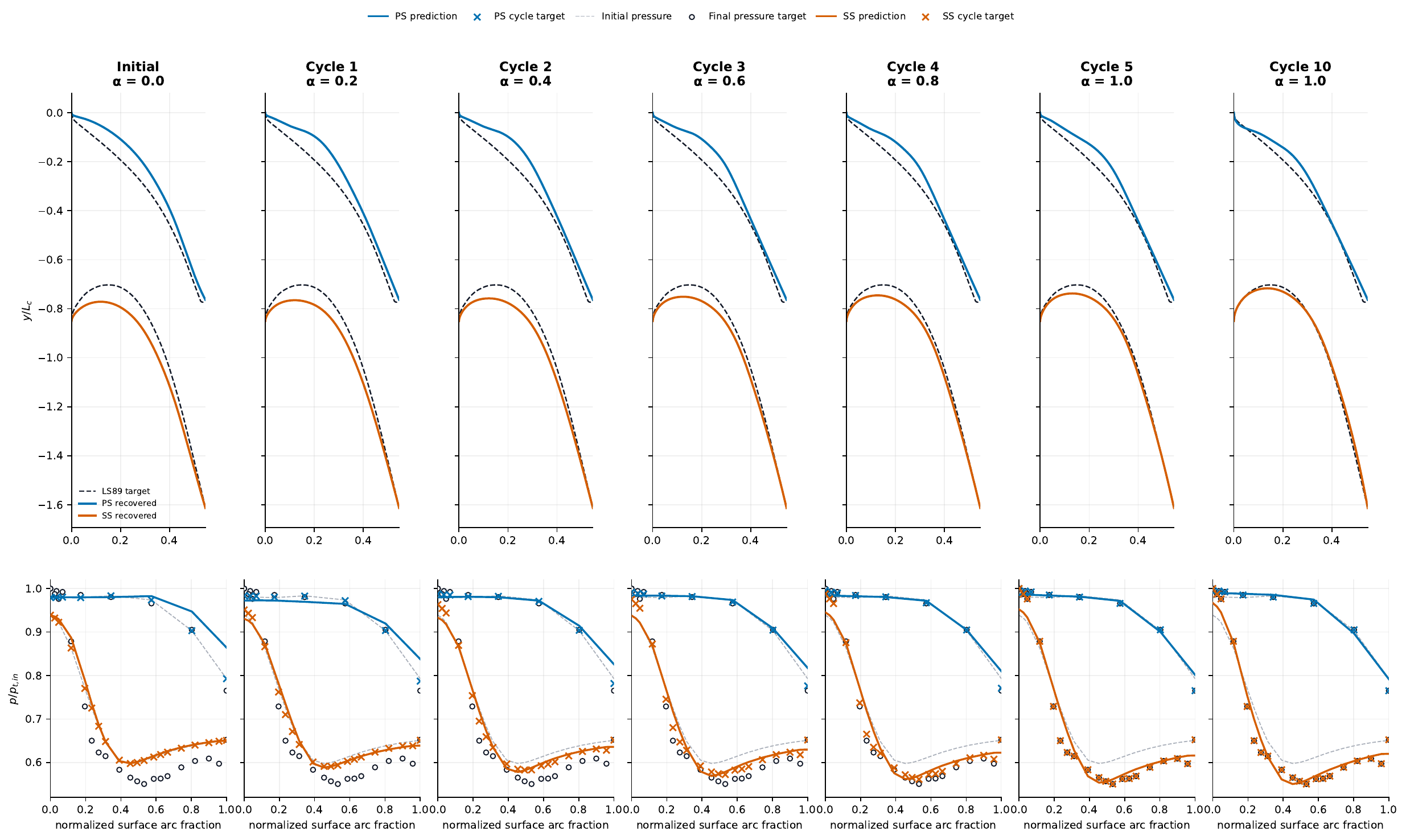}
\caption{Initial state, Cycles 1--5, and Cycle 10. Crosses indicate scheduled targets; open circles indicate the final prescribed pressure target $p^\star$. Each column pairs geometry and flow from its own state. Cycle-end errors against the final target are given in Table~\ref{tab:cycles}.}\label{fig:ramp}
\end{figure}

Table~\ref{tab:cycles} tracks pressure error against the same final target throughout training. Figure~\ref{fig:ramp} pairs the evolving geometry with the pressure distribution and the intermediate target prescribed in each cycle. During the five-cycle pressure ramp, the mean final-target pressure error decreases from 3.8077\% at Cycle 1 to 2.2042\% at Cycle 5. The first cycle mainly changes the suction surface; pressure-side recovery becomes more pronounced in subsequent cycles.
\begin{table}[pos=htbp]
\caption{Cycle-end pressure errors against the final target, including during the homotopy ramp.}\label{tab:cycles}
\centering\begin{tabular}{rrr}\toprule
Cycle&$\alpha$&$\bar E_p$ (\%)\\\midrule
1&0.2&3.8077\\
2&0.4&3.1708\\
3&0.6&2.7891\\
4&0.8&2.4925\\
5&1.0&2.2042\\
6&1.0&2.0502\\
7&1.0&1.8921\\
8&1.0&2.0008\\
9&1.0&1.6970\\
10&1.0&1.6682\\\bottomrule
\end{tabular}\end{table}

After the full target is reached, mean pressure error decreases further to 1.6682\% at Cycle 10, with a temporary increase at Cycle 8. The late-cycle improvement becomes small: the mean error decreases from 1.6970\% at Cycle 9 to 1.6682\% at Cycle 10, a reduction of 0.0288 percentage points. In Fig.~\ref{fig:pressure}(b), the target points generally lie on both sides of the predicted pressure curves, with local deviations remaining. These observations support Cycle 10, the endpoint of the prescribed training budget, as a practical stopping point. Further reduction of the pressure-fitting error alone was not taken as sufficient reason to continue modifying the geometry. The pressure response of its redesigned geometry is then assessed by independent forward CFD in Section~\ref{sec:forward_results}.

\subsection{Comparison with direct joint PINN optimisation}
\label{sec:direct_joint}
\begin{figure}[pos=htbp]
\centering\includegraphics[width=0.95\linewidth]{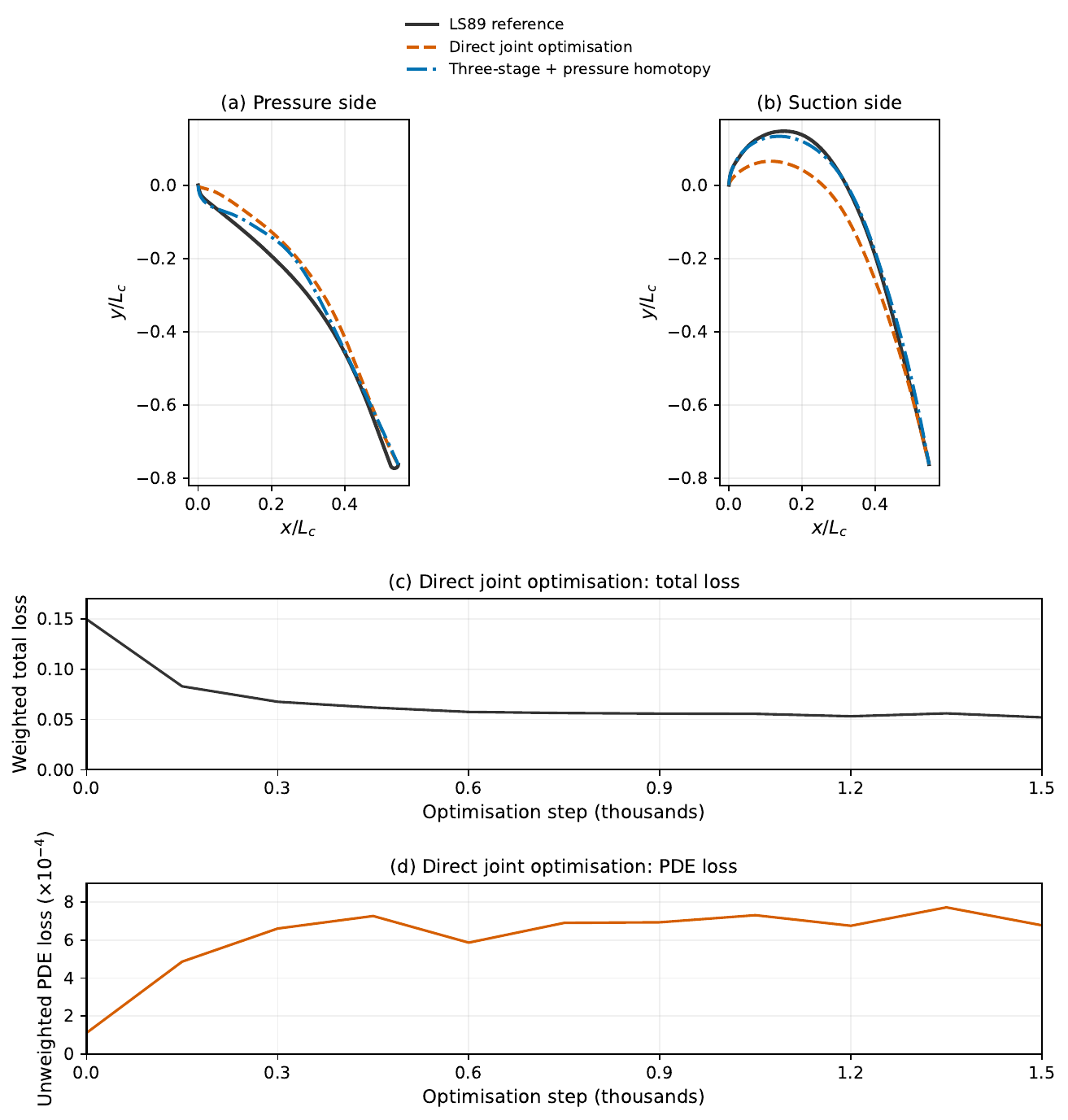}
\caption{Direct joint PINN optimisation compared with the three-stage pressure-homotopy result. (a,b) Final geometries against the LS89 reference. (c,d) Weighted total loss and unweighted PDE loss of the direct run, using a fixed target and fixed weights. Lines connect all eleven recorded samples without smoothing.}\label{fig:direct_joint}
\end{figure}

Figure~\ref{fig:direct_joint} compares a direct joint PINN result with the three-stage pressure-homotopy result. The direct method applies the final pressure target from the outset and updates the flow and geometry simultaneously. Its loss weights remain fixed throughout optimisation.

The direct PINN achieves pressure-side and suction-side pressure RMSEs of 1.61\% and 2.04\%, respectively, while its recovered profile remains visibly different from the reference. The three-stage result more closely follows the reference outline, particularly on the suction surface. During the direct run, the weighted total loss decreases from 0.1496 to 0.0521 and the pressure loss from 0.00411 to 0.000675, whereas the unweighted PDE residual increases overall from $1.11\times10^{-4}$ to $6.78\times10^{-4}$. Thus, a lower total loss and a good pressure fit can coexist with poorer satisfaction of the governing equations and limited shape recovery.

This example illustrates the competition between pressure fitting and the physics constraints in direct joint optimisation. The optimizer reduces the composite objective while accepting a larger PDE residual. It motivates coordinating pressure fitting, boundary motion and subsequent flow re-adaptation, and evaluating the resulting blade with independent forward CFD.

\subsection{Equation residuals and boundary consistency}

The mean sampled PDE loss is $3.7050\times10^{-4}$ with standard deviation $9.4529\times10^{-6}$ across five point sets, a relative sampling spread of about 2.55\%. Table~\ref{tab:physics} also reports wall and periodic losses. Periodic mismatch remains approximately $2.16\times10^{-3}$ in its implemented squared-loss measure.
\begin{table}[pos=htbp]
\caption{Frozen-model diagnostics across five independently sampled evaluation point sets. Means and standard deviations use the nondimensional loss definitions.}\label{tab:physics}
\centering\begin{tabular}{lrr}\toprule
Diagnostic & Mean & Standard deviation\\\midrule
Euler residual loss &$3.7050\times10^{-4}$&$9.4529\times10^{-6}$\\
Wall loss &$3.1376\times10^{-4}$&$2.8379\times10^{-5}$\\
Periodic loss &$2.1645\times10^{-3}$&$1.3058\times10^{-4}$\\
Inlet loss &$5.8105\times10^{-4}$&$5.7844\times10^{-5}$\\
Outlet loss &$1.0770\times10^{-4}$&$7.5458\times10^{-6}$\\
Total-temperature loss &$1.4914\times10^{-4}$&$7.0151\times10^{-6}$\\
\bottomrule
\end{tabular}\end{table}
\FloatBarrier
These residual and boundary diagnostics are complemented by the independent forward pressure assessment below.

\FloatBarrier
\subsection{Independent forward assessment of the recovered blade}
\label{sec:forward_results}

\begin{figure}[pos=htbp]
\centering\includegraphics[width=0.98\linewidth]{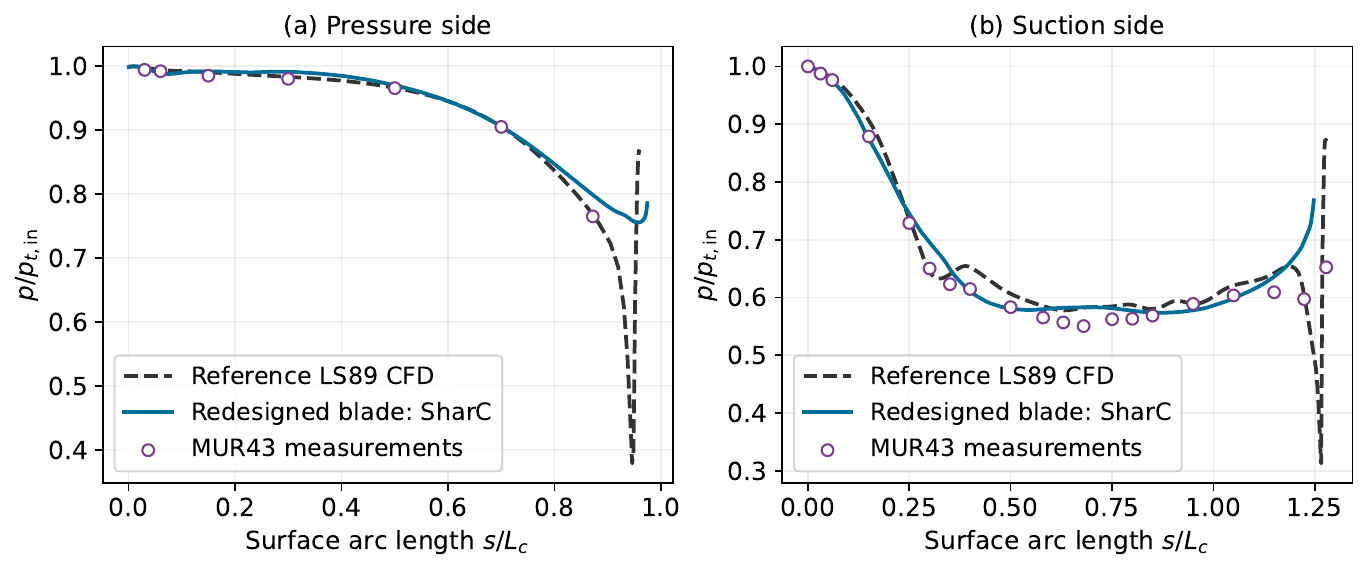}
\caption{Independent forward pressure comparison for the Cycle 10 recovered blade. (a) Pressure side and (b) suction side: recovered-blade SharC pressures (solid), published reference-LS89 CFD (dashed), and MUR43 experimental measurements (circles). The reference CFD curve was digitized from Fig.~13 of Post et al.\ \cite{internal}. Each surface uses its own leading-edge arc-length coordinate.}\label{fig:forward_cfd}
\end{figure}

Figure~\ref{fig:forward_cfd} compares the SharC pressure distribution on the Cycle 10 recovered blade with the MUR43 experimental measurements and the published reference-LS89 CFD curve of Post et al.\ \cite{internal}. The forward calculation uses the Cycle 10 geometry shown in Fig.~\ref{fig:geometry}. This comparison tests whether the modified boundary produces the intended loading in a separate flow calculation.
\FloatBarrier

The recovered-blade calculation reproduces the main loading trends on both surfaces in Fig.~\ref{fig:forward_cfd}. It captures the gradual pressure decrease on the pressure side and the initial acceleration, broad low-pressure region and subsequent recovery on the suction side. The largest local differences occur near the trailing edge, where the sharp pressure variation in the published reference curve is not reproduced in the same form. The agreement with the experimental loading in this viscous forward calculation supports the use of Euler physics for pressure-driven redesign at the selected operating point.

We attribute the remaining trailing-edge differences to two aspects of the geometry and pressure-data representation. First, the parameterisation explicitly treats the leading-edge connection but assumes that the two surfaces meet at a sharp trailing edge. It therefore does not reproduce the rounded termination of the physical LS89 blade or the way the pressure-side and suction-side streams leave that finite-radius edge, which affects the local pressure response. Second, the available experimental pressure points are too sparse near the trailing edge to constrain its rapid local pressure variation in detail. The inverse update is therefore weakly informed about both the local edge curvature and the associated pressure response. Denser pressure measurements in this region, additional local geometry controls and an explicit trailing-edge rounding parameter provide direct avenues for improving this part of the reconstruction.

\subsection{Computational resources and runtime}
\label{sec:runtime}

The calculations were performed under Ubuntu on a CPU using eight PyTorch threads. The three-hidden-layer network of 20 units per layer contains 984 trainable parameters, with a further 16 trainable geometry ordinates. The complete ten-cycle, 18,000-step inverse-training process takes 821.5~s, or approximately 13.7~min; independent forward CFD verification is excluded. For comparison, a viscous compressible SharC verification calculation using the same finest structured mesh of 8,960 cells takes approximately 4~min in the same CPU environment. The complete PINN inverse-training time therefore corresponds to roughly three to four such forward CFD calculations. The compact network and low-dimensional geometry representation allow repeated flow adaptation and boundary updates to be completed in a short time even on a CPU.

\FloatBarrier
\section{Discussion}
\subsection{Interpretation of the pressure--flow--geometry coupling}
The method addresses modification of an initial blade to meet a prescribed pressure-loading objective. Pressure homotopy defines intermediate design problems, the frozen-flow geometry stage moves the walls using pressure and tangency information, and re-anchoring adapts the state to each changed boundary. The LS89 experiment demonstrates these operations in a controlled reference-based setting. The primary result is target-pressure matching with an explicitly modified geometry; the reference blade supplies a visual comparison, and independent forward CFD tests the pressure response of the resulting design.

The LS89 trajectory demonstrates the feasibility of the combined procedure. Curriculum and continuation literature motivate the pressure schedule, while state--design iteration motivates the staged coupling. The direct joint example in Section~\ref{sec:direct_joint} shows that decreasing total and pressure losses can accompany an increasing PDE residual. It illustrates why pressure fit and the learned physical state should be examined together when interpreting the resulting geometry. For industrial blade design, final independent forward CFD verification remains essential to assess the pressure response and physical behaviour of the modified blade.

\subsection{Role within the blade-design process}

The formulation targets a common engineering task after mean-line design has established the principal stage requirements and operating point. At this stage, the inlet total state and flow direction, outlet pressure, cascade pitch and other boundary specifications are available, while the detailed blade section must be shaped to deliver a prescribed loading. The present method uses these operating conditions and a target surface-pressure distribution to update the blade surfaces. This provides a route from an aerodynamic loading specification to a modified section geometry within the detailed-design process \cite{aungier2006,li2017turbomachinery}.

In the LS89 turbine-cascade case, the pressure targets originate from experimental measurements, and the resulting blade is assessed by an independent forward CFD calculation. The recovered pressure trends demonstrate the feasibility of using this procedure to modify a blade under specified operating conditions. In an engineering application, pressure targets can be selected to adjust the loading distribution, while geometric constraints can encode required endpoint locations, clearances and thickness limits. Forward CFD then closes the design loop by checking the pressure response of the modified blade before further performance assessment.

\subsection{Physical-model scope and future extensions}

The present assessment uses a reduced-thickness initial blade with the reference camber line and a flow network pretrained on CFD data for that initial geometry. Sensitivity to alternative starting shapes and to the accuracy of the preliminary flow fit remains to be assessed. The pressure targets were digitized from a published experimental plot without an accompanying measurement-uncertainty model, so the reported fitting errors do not quantify uncertainty in the inferred design.

The inverse model uses the steady compressible Euler equations and does not explicitly represent viscosity, boundary-layer transition or viscous separation. The agreement obtained with experimental-origin pressure data and independent SharC verification supports its use for the present loading-driven redesign, while its applicability to flows dominated by viscous effects remains to be examined. Low-speed or low-Reynolds-number turbine flows with strong boundary-layer growth, transition or separation require a viscous formulation and appropriate near-wall resolution. A steady Navier--Stokes or suitable turbulence-model extension would also enable assessment of aerodynamic losses and heat transfer.

Transonic turbine conditions with shocks form another important test. Although the Euler equations admit shock solutions, the present smooth-network, primitive-residual formulation has not been verified for shock-containing flows. Shock resolution and, where relevant, shock--boundary-layer interaction require additional numerical treatment and validation. Further work should also examine different operating points and loading targets, refine the trailing-edge representation and pressure sampling, and assess mesh sensitivity in the independent CFD verification. These extensions would establish the operating range over which the present design procedure can be applied reliably.

Extension to three-dimensional blades is another priority, since spanwise variations in loading and geometry, secondary flows and endwall effects are integral to practical blade design. One route is a fully three-dimensional compressible PINN with a differentiable blade-surface representation and joint treatment of the three velocity components. An alternative is a spanwise-coupled sectional formulation: multiple two-dimensional sections are linked through spanwise velocity and transport or source terms derived from the three-dimensional conservation equations. Such coupling must preserve cross-section compatibility and account for transport in the spanwise direction. These two routes, together with a Navier--Stokes extension and tests on additional blade geometries and operating conditions, provide a development path towards three-dimensional viscous compressible redesign.

\section{Conclusions}

A PINN-based method has been developed for modifying turbine-blade geometry to match a prescribed surface-pressure distribution. A differentiable blade representation couples the flow and geometry within one optimisation framework. Pressure-target continuation and the three-stage sequence of flow adaptation, frozen-flow geometry updates and flow re-anchoring coordinate the evolution of the pressure field and moving boundaries.

For the LS89 case, the final pressure-side and suction-side fitting errors are 1.128\% and 2.209\% of inlet total pressure, with a surface-balanced mean of 1.668\%. Although the redesigned blade differs from the reference outline, independent SharC calculations reproduce similar principal surface-pressure trends. This is consistent with the non-uniqueness of pressure-based inverse design: geometrically different blades can produce similar loading under the prescribed operating conditions. The forward assessment therefore supports the aerodynamic relevance of the recovered shape beyond its resemblance to the reference geometry.

The differentiable formulation offers substantial potential for reducing redesign cost. Automatic differentiation supplies update directions for the flow and geometry within the learned representation, reducing reliance on repeated external CFD solves and separate CFD sensitivity calculations during the inverse iteration. These gradients provide a quantitative, traceable link between geometric changes and the pressure and constraint objectives, supporting systematic and reproducible design updates. The compact implementation completes the present inverse-training process in approximately 13.7~min on a CPU. Independent forward CFD remains an essential final verification step in industrial use.

Future development will extend the formulation to viscous compressible flow and three-dimensional blade design, through fully three-dimensional PINNs or spanwise-coupled sectional models. Broader geometry and operating-condition studies, together with improved trailing-edge representation and pressure sampling, will further establish its engineering applicability.

\section*{CRediT authorship contribution statement}
\textbf{Yuan Huang:} Conceptualization, Methodology, Software, Investigation, Validation, Formal analysis, Visualization, Writing -- original draft, Writing -- review \& editing. \textbf{Francesca di Mare:} Conceptualization, Supervision, Project administration, Funding acquisition, Writing -- review \& editing.
\section*{Declaration of competing interest}
The authors declare that they have no known competing financial interests or personal relationships that could have appeared to influence the work reported in this paper.
\section*{Funding}
This work was funded by the Deutsche Forschungsgemeinschaft (DFG, German Research Foundation) within Priority Programme SPP 2403.
\section*{Data availability}
The data supporting the findings of this study are available from the corresponding author upon reasonable request.
\section*{Declaration of generative AI and AI-assisted technologies in the manuscript preparation process}
During the preparation of this manuscript, OpenAI Codex was used to assist with drafting and revising the manuscript text and title, improving the wording and clarity of presentation, checking LaTeX equations, and verifying references. The authors wrote the initial draft, conducted the experiments, and reviewed and edited the manuscript. The authors take full responsibility for its scientific content.

\appendix
\section{Detailed auxiliary loss definitions}
\label{app:loss_definitions}

With stored raw surface labels $s_{s,i}$, let $\Delta s_{s,i}=s_{s,i+1}-s_{s,i}$, let $\hat p_{s,i}$ denote predicted pressure, and define
\begin{equation}
 G_s^{(c)}=\left[\frac{1}{N_s-1}\sum_{i=1}^{N_s-1}
 \left(\frac{p^{(c)}_{s,i+1}-p^{(c)}_{s,i}}{\Delta s_{s,i}}\right)^2\right]^{1/2}.
\end{equation}
The geometry-stage gradient penalty is
\begin{equation}
 \mathcal L_{\nabla p}^{(c)}=\sum_s\frac{1}{N_s-1}\sum_{i=1}^{N_s-1}
 \left[\frac{(\hat p_{s,i+1}-\hat p_{s,i})/\Delta s_{s,i}-(p^{(c)}_{s,i+1}-p^{(c)}_{s,i})/\Delta s_{s,i}}
 {\max(G_s^{(c)},10^{-6})}\right]^2.
\end{equation}
Its weight is 0.01 only during geometry updates, and its reference is the current homotopy target.

For wall sample $k$ on side $s$, let $v_{n,s,k}=\mathbf u_{s,k}\!\cdot\!\mathbf n_{s,k}$ and let $\xi_{s,k}$ be fractional arc distance from the leading edge. With detached weights
\begin{equation}
 \tilde w_{s,k}=1+18\exp[-(\xi_{s,k}/0.15)^2],\qquad
 w_{s,k}=\frac{\tilde w_{s,k}}{N_w^{-1}\sum_k\tilde w_{s,k}},
\end{equation}
the implemented wall losses are
\begin{align}
 \mathcal L_w&=\sum_s\frac{1}{N_w}\sum_{k=1}^{N_w}w_{s,k}v_{n,s,k}^2,\\
 \mathcal L_{\rm tan}&=\sum_s\frac{\sum_{k=1}^{N_w}a_{s,k}
 [v_{n,s,k}/\max(\|\mathbf u_{s,k}\|,10^{-6})]^2}
 {\max(\sum_{k=1}^{N_w}a_{s,k},1)},
\end{align}
where the detached mask is $a_{s,k}=\mathbf 1[\|\mathbf u_{s,k}\|\ge0.02\max_j\|\mathbf u_{s,j}\|]$. The tangency weight is 0.5 in every stage.

The geometry regularizers use 256 uniformly spaced $x_m$ locations. With passage clearance $g_m=y_\PS(x_m)-y_\Ssurf(x_m)$,
\begin{align}
 \mathcal L_{\rm gap}&=\frac1{256}\sum_{m=1}^{256}[\max(0,0.08-g_m)]^2,\\
 \mathcal L_{\rm bounds}&=\frac1{256}\sum_{m=1}^{256}\bigl(
 [\max(0,y_\PS(x_m)-0.12)]^2+[
 \max(0,-0.90-y_\PS(x_m))]^2\nonumber\\
 &\hspace{31mm}+[\max(0,y_\Ssurf(x_m)+0.50)]^2+[
 \max(0,-1.72-y_\Ssurf(x_m))]^2\bigr),\\
 \mathcal L_{\rm smooth}&=\sum_{s\in\{\PS,\Ssurf\}}\frac16\sum_{k=1}^{6}
 (q_{s,k+2}-2q_{s,k+1}+q_{s,k})^2.
\end{align}
Here $\mathcal L_{\rm gap}$ constrains passage clearance rather than blade thickness. The smoothness term acts on the eight interior ordinates on each side, omits endpoints, and contains no division by station spacing.

\bibliographystyle{elsarticle-num}
\bibliography{references}
\end{document}